\documentclass{egpubl}
\usepackage{3dor16}
\usepackage{amssymb, amsmath}
\newtheorem{thm}{Theorem}[section]

\numberwithin{equation}{subsection}

\newcommand{\R}{{\mathbb{R}}}
\newcommand{\Homeo}{\mathrm{Homeo}}

\newcommand{\p}{{\varphi}}

\WsSubmission    
 \electronicVersion 

\ifpdf \usepackage[pdftex]{graphicx} \pdfcompresslevel=9
\else \usepackage[dvips]{graphicx} \fi

\PrintedOrElectronic

\usepackage{t1enc,dfadobe}

\usepackage{egweblnk}
\usepackage{cite}

\title[Towards an observer-oriented theory of shape comparison]%
      {Position paper:\\Towards an observer-oriented theory of shape comparison}

\author[P. Frosini]
       {P. Frosini
        \\
         University of Bologna, Italy\\
         ARCES, Bologna, Italy
       }

\begin{document}


\maketitle

\begin{abstract}
In this position paper we suggest a possible metric approach to shape comparison that is based on a mathematical formalization of the concept of observer, seen as a collection of suitable operators acting on a metric space of functions. These functions represent the set of data that are accessible to the observer, while the operators describe the way the observer elaborates the data and enclose the invariance that he/she associates with them. 
We expose this model and illustrate some theoretical reasons that justify its possible use for shape comparison.


\end{abstract}

\section{Introduction}

The poem ``The Blind Men and the Elephant'' by John Godfrey Saxe begins with these verses:

\begin{verse}\em
``It was six men of Indostan \\
To learning much inclined, \\
Who went to see the Elephant \\
(Though all of them were blind), \\
That each by observation \\
Might satisfy his mind.''
\end{verse}

This famous Indian story describes a group of blind men that touch an elephant and totally disagree about what it is like, because each one touches a different part of the body. In literature, art and philosophy the theme of multiperspectivity (i.e., the existence of many different interpretations of the same perceptual phenomenon) is a well-known issue \cite{Mau11}.

Beautiful examples of this fact can also be found in artworks. The fascinating sculptures by Guido Moretti displayed in Figures 1 and 2 show in a direct and clear way how the concept of shape cannot be defined and treated independently from the choice of an observer. 
Shape is indeed in the beholder's eyes, and phenomena such as camouflage and optical illusions depend on this basic principle (cf., e.g., \cite{Koe90,Fro09,GFS*10}). 

It is indeed well-known that in many contexts the concept of shape is not a property of objects but of the pairs (object, observer) that are involved in perception, since changing the observer can drastically transform the perception of reality. 

In the past these observations were mostly confined to the philosophical and epistemological debate, but nowadays they start to be quite relevant also in several scientific applications involving Information Technology \cite{Sta07}. 
In particular, geometrical shape comparison often requires approaches that take into account the role of the observer. 
Trying to avoid this problem by considering shapes just as subsets, topological subspaces or submanifolds of an Euclidean space contributes to the semantic gap between the geometrical descriptions and their perceptual meanings (cf. \cite{HF07,SWS*00}).
This semantic gap cannot be closed focusing just on the objects and disregarding the chosen observer.

In this position paper we intend to consider this issue and propose a possible solution, framing it in the emerging field of Topological Data Analysis \cite{Car09}.

We are interested in these questions:
\begin{itemize}
\item Is there a general metric model to compare data in TDA?
\item What is the role of the observer in this comparison?
\item How could we approximate the observer's judgement by means of a computable metric?
\end{itemize}

\begin{figure}[htb]
  \centering
  \includegraphics[width=.8\linewidth]{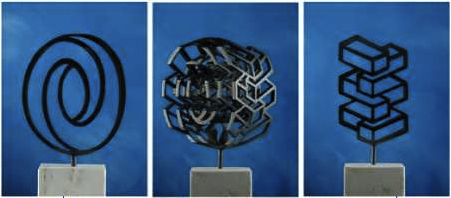}
   \parbox[t]{.9\columnwidth}{\relax
     }
  \caption{\label{fig1} Different observers can perceive different shapes in the presence of the same object. 
  This image depicts three views of the bronze sculpture ``Impossible Ring and Parallelepipeds'' by Guido
Moretti.}
\end{figure}

\begin{figure}[htb]
  \centering
  \includegraphics[width=.8\linewidth]{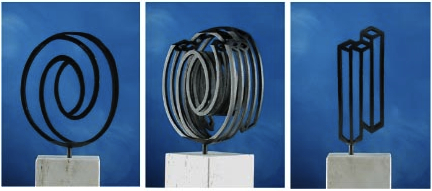}
   \parbox[t]{.9\columnwidth}{\relax
     }
  \caption{\label{fig2} Three views of the bronze sculpture ``Impossible Ring and Pillars'' by Guido
Moretti.}
\end{figure}

\section{Our theoretical model}

In the next two sections we will describe both the principles accounting for our model and its mathematical formalization.

\subsection{An informal description of our model}




The model we propose to consider is based on these general assumptions:
\begin{enumerate}
\item \label{a} No object can be studied in a direct and absolute way. Any object is only knowable through acts of measurement made by an observer.
\item \label{b} Any act of measurement can be represented as a function defined on a topological space.
\item \label{c} The observer usually acquires measurement data by applying operators to the functions describing them.
\item \label{d} Only the observer is entitled to decide about shape similarity.
\end{enumerate}  

Assumption \ref{a} is justified by the fact that according to the scientific paradigm, we cannot refer to properties of reality that are intrinsically impossible to detect by measurement processes. 

Assumption \ref{b} is based on the fact that when we make a measurement, we usually obtain a function as a result. 
For example, a grayscale image can be considered as a function from a rectangle to the real numbers (or, in the discrete case, from the set of cells of a matrix to a set of integers). The result of a CT scanning can be seen as a function from $S^1$ (or, more precisely, a helix going around a body) to the real numbers, where $S^1$ represents the topological space of all directions that are orthogonal to a given axis and the real numbers are the metric space of all possible quantities of matter encountered by the X-ray beam in the considered direction. A weight measurement is just a function from a singleton to the real numbers, taking the only available point in the domain to the weight of the object we are examining. We also observe that many kinds of data that are not usually represented as functions can in fact be described by means of functions. For example, every cloud $C$ of points in a metric space $M$ is equivalent to the function that takes each point of $M$ to its distance from $C$.

The requirement that the domain of the function be a topological space is important in applications where we need to assume that our data are continuous. However, in the presence of discontinuities, it is usually required that they be localized at ``small'' subsets of the domain, so that we still need to use a topology on it. For example, we usually assume that the color of the points of an object 
changes continuously, possibly apart from a set of null measure. The formalization of this assumption cannot be made without the use of a topology.

Assumption \ref{c} is supported by the fact that in most of the experiments, data are not used directly, but after an elaboration 
that makes easier (or simply feasible) their analysis. This elaboration is usually done by means of suitable operators, which are sometimes 
embedded in the measurement process. Building the body of a simplicial $n$-complex from a cloud of points or blurring an image are examples of two such operators among many possible others. These operators transform functions into other functions that are usually simpler to manage.

It is important to underline that the observer cannot usually choose the functions representing the measurement data, but can often choose the operators that will be applied to those functions. Moreover, the choice of the operators reflects the invariances that are relevant for the observer.




Assumption \ref{d} is based on what we previously said about multiperspectivity.

According to this model, instead of directly focusing on the objects we are interested in, we should focus on the functions describing the measurements we make on the objects, and on the ``glasses'' that we use to ``observe'' the functions. 
In our approach, these ``glasses'' are invariant operators which act on the functions. 

These operators represent the observer's perspective and, in some sense, define the observer.

\begin{figure}[htb]
  \centering
  \includegraphics[width=.8\linewidth]{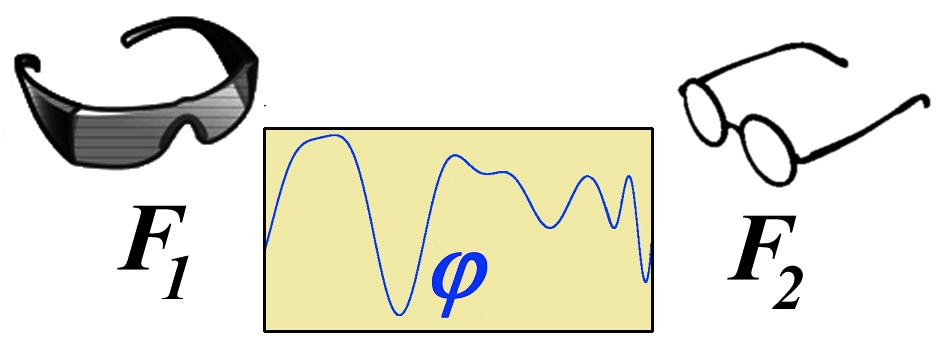}
   \parbox[t]{.9\columnwidth}{\relax
     }
  \caption{\label{fig4} In the proposed model, each observer can be represented as a collection of (suitable) operators $F_i$, 
  which act on the functions that represent the measurement data and are endowed with the invariance that the observer has chosen.}
\end{figure}

\subsection{A mathematical framework to formalize our model}

The previous epistemological model has led to the mathematical framework that will be described in this section (cf. \cite{FJ16}).

Let us consider a compact space $X$ and a subset $\Phi$ of the set $C^0(X,\R^k)$ of all continuous functions from $X$ to $\R^k$. 
The space $\Phi$ represents the space of the functions that the observer considers as acceptable data. 
In our previous example about CT scanning,  $\Phi$ would be the set of all functions that we can obtain by associating each X-ray beam with the quantity of matter it can encounter.

The observer usually takes some invariance into account. This invariance can be usually represented by a group.

Let $G$ be a subgroup of the group $\mathrm{Homeo}(X)$ of all homeomorphisms $f:X\to X$. 
We assume that the group $G$ acts on $\Phi$ by composition on the right, i.e. by taking every $\varphi\in\Phi$ to the function $\varphi\circ g\in\Phi$, for each $g\in G$.

We can define a pseudo-distance $d_G:\Phi\times \Phi\to\R$ by setting 
$$d_G(\varphi_1,\varphi_2)=\inf_{g \in G}\max_{x \in X}\|\varphi_1(x)-\varphi_2(g(x))\|.$$
The function $d_G$ is called the {\em natural pseudo-distance associated with the group $G$}.
We recall that a pseudo-distance is just a distance $d$ without the property $d(a,b)=0\implies a=b$.

In plain words, the definition of $d_G$ is based on the attempt of finding the best correspondence between the functions (i.e. observations) $\varphi_1,\varphi_2$ by means of homeomorphisms in $G$. If $d_G(\varphi_1,\varphi_2)=0$, $\varphi_1$ and $\varphi_2$ are equivalent w.r.t. $G$. For example, if $\Phi$ is the space of all normalized grayscale images, represented as the set of all compact-supported functions from the real plane to the interval $[0,1]$, we can choose $G$ to be the group of rigid motions of the plane. In this case the equality $d_G(\varphi_1,\varphi_2)=0$ means that there is a rigid motion taking the image $\varphi_1$ to the image $\varphi_2$. The choice of the invariance group $G$ is assigned to the observer, which is the only judge of similarity in shape comparison.


The natural pseudo-distance $d_G$ represents our ground truth. When the observer has chosen the set $\Phi$ of signals he/she can perceive and the invariance group $G$ he/she uses to define which signals are considered equivalent, $d_G$ endows $\Phi$ with a pseudo-metric structure. 
Unfortunately, in many cases $d_G$ is difficult to compute. This is also a consequence of the fact that we can easily find subgroups $G$ of $\Homeo(X)$ that cannot be approximated with arbitrary precision by smaller finite subgroups of $G$ (i.e. $G=$ group of rigid motions of $X=\R^3$).

Nevertheless,  $d_G$ can be approximated with arbitrary precision by means of a dual approach based on persistent homology and $G$-invariant non-expansive operators.

We recall that persistent homology is a theory describing the $m$-dimensional holes (components, tunnels, voids, ... ) of the sublevel sets of a topological space $X$ endowed with a continuous 
function $\varphi:X\to\R^k$. In the case $k=1$, persistent homology is described by suitable collections of points called \emph{persistence diagrams} \cite{EH10}. These diagrams can be compared by a suitable metric $d_{\mathrm{match}}$, called \emph{bottleneck} (or \emph{matching}) \emph{distance}. The simplest version of this theory counts the components of the sub-level sets of $\varphi$ \cite{VUF*93}. 

For the sake of simplicity, in the rest of this paper we will assume that $k=1$.
For technical reasons, let us also assume that 
the topological space $X$ is finitely triangulable and has nontrivial homology in degree $m$, and that the set $\Phi$ contains the set of all constant functions.

Now, let us consider the set $\mathcal{F}^\mathrm{all}(\Phi,G)$ of all $G$-invariant non-expansive operators (GINOs) from $\Phi$ to $\Phi$.

In plain words, $F\in \mathcal{F}^\mathrm{all}(\Phi,G)$ means that
\begin{enumerate}
\item $F:\Phi\to \Phi$
\item $F(\varphi\circ g)=F(\varphi)\circ g$. ($F$ is a $G$-operator)
\item $\|F(\varphi_1)-F(\varphi_2)\|_\infty\le \|\varphi_1-\varphi_2\|_\infty$. ($F$ is non-expansive)
\end{enumerate}

The symbol $\|\cdot\|_\infty$ denotes the sup-norm.


In the example where $\Phi$ is the space of all normalized grayscale images and $G$ is the group of rigid motions of the plane, a simple example of operator $F\in \mathcal{F}^\mathrm{all}(\Phi,G)$ is given by the 
Gaussian blurring filter, i.e. the operator $F$ taking $\varphi\in\Phi$ to the function 
$\psi(x)=\frac{1}{2\pi\sigma^2}\int_{\R^2}\varphi(y)e^{-\frac{\|x-y\|^2}{2\sigma^2}}\ dy$.

Now, let us assume that $\mathcal{F}$ is a subset of $\mathcal{F}^\mathrm{all}(\Phi,G)$. For every $\varphi_1,\varphi_2\in \Phi$ we can consider the supremum $D^\mathcal{F}_{\mathrm{match}}(\varphi_1,\varphi_2)$ of the bottleneck distances between the persistence diagrams (in the fixed degree $m$) of the functions  $F(\varphi_1),F(\varphi_2)$, when $F$ varies in $\mathcal{F}$.


Since $D^\mathcal{F}_{\mathrm{match}}$ is the supremum of a set of pseudo-metrics, it is itself a pseudo-metric.
Furthermore, 
for every $\p_1,\p_2\in \Phi$ and every $g\in G$ the equalities $D^\mathcal{F}_{\mathrm{match}}(\p_1\circ g,\p_2)=D^\mathcal{F}_{\mathrm{match}}(\p_1,\p_2\circ g)=D^\mathcal{F}_{\mathrm{match}}(\p_1,\p_2)$ hold.

We remark that the pseudo-distance $D^\mathcal{F}_{\mathrm{match}}$ and the natural pseudo-distance $d_G$ are  defined in quite different ways. 
In spite of this, the following result can be proved \cite{FJ16}.

\begin{thm}
If $\mathcal{F}=\mathcal{F}^\mathrm{all}(\Phi,G)$, then the pseudo-distance $D^\mathcal{F}_{\mathrm{match}}$ coincides with the natural pseudo-distance $d_G$.
\end{thm}

This fact suggests to study $D^\mathcal{F}_{\mathrm{match}}$ instead of $d_G$.


We can prove that if $\Phi$ is a compact metric space with respect to the sup-norm, then $\mathcal{F}^\mathrm{all}(\Phi,G)$ is a compact metric space with respect to the distance $d$ defined by setting $$d(F_1,F_2):=\max_{\p\in \Phi}\|F_1(\p)-F_2(\p)\|_\infty$$ for every $F_1,F_2\in \mathcal{F}$ \cite{FJ16}.

As a consequence, we can also prove that if the metric space $\Phi$  is compact with respect to the sup-norm and $\mathcal{F}$ is a subset of $\mathcal{F}^\mathrm{all}(\Phi,G)$, then 
for every $\epsilon >0$ a finite subset $\mathcal{F}^*$ of $\mathcal{F}$ exists, such that $$\left|D^{\mathcal{F}^*}_{\mathrm{match}}(\varphi_1,\varphi_2)-D^\mathcal{F}_{\mathrm{match}}(\varphi_1,\varphi_2)\right|\le \epsilon$$
for every $\p_1,\p_2\in \Phi$.

This statement implies that the pseudo-distance $D^\mathcal{F}_{\mathrm{match}}$ (and hence also $d_G$) can be approximated computationally, at least in the case that $\Phi$ is compact. As we have just seen, this is done by means of a collection $\mathcal{F}^*$ of suitable operators, which takes the place of the observer in our model.

It is important to highlight that in the framework we have described the invariance group $G$ is a variable of our problem, and that its choice is completely assigned to the observer, according to the statement that only the observer is entitled to decide about shape similarity. 

Many interesting questions remain open. The most important is probably the one of devising methods to build families $\mathcal{F}^*$ of $G$-invariant non-expansive operators that are small and simple to compute, but still able to guarantee that the associated pseudo-metric $D^{\mathcal{F}^*}_{\mathrm{match}}$ is a good approximation of the natural pseudo-distance $d_G$.
\subsection{A simple case study in this model}

In order to show the use of our approach, we have realized (jointly with Grzegorz Jab{\l}o\'nski and Marc Ethier) a simple demonstrator that illustrates how our model based on collections of $G$-invariant non-expansive operators (GINOs) could make available new methods for image comparison. The demonstrator (named GIPHOD--\emph{$G$-Invariant Persistent HOmology Demonstrator}) is available at the web page \url{http://giphod.ii.uj.edu.pl/}. The program asks the user to choose an invariance group in a list and a query image in a dataset $\Phi^*$ of quite simple synthetic images obtained by adding a small number of bell-like functions. After that, GIPHOD provides ten images that are judged to be the most similar to the proposed query image with respect to the chosen invariance group. In this case study, the dataset $\Phi^*$ is a subset of the set $\Phi$ of all continuous functions from the square $[0,1]\times[0,1]$ to the interval $[0,1]$. Each of them represents a grayscale image on the square $[0,1]\times[0,1]$ ($1$=white, $0$=black).

GIPHOD works by using a collection of GINOs for each invariance group $G$. This demonstrator tries to approximate $d_G$ by means of the previously described technique, based on the persistent homology of the functions $F(\varphi)$ for $\varphi\in\Phi$ and $F$ varying in our set of operators.

\subsection{Conclusions}

The model that we have proposed to study is based on the idea that, from the mathematical point of view, a shape should not be considered as a subset or a submanifold of a Euclidean space, but as a quotient of the space $\Phi$ of the signals that can be perceived by the chosen observer with respect to the action of a given invariance group $G$. According to this model, each observer should be represented by a collections of group-invariant non-expansive operators acting on $\Phi$. This idea is supported by some formal results showing how the emerging theory of persistent homology could be used to study the approach to shape comparison that we have proposed in this position paper. We suggest that this approach could possibly contribute to bridge the semantic gap by means of the framework of topological data analysis.

Developments of the proposed model are presently the object of research. In particular, the extension of the model to the case of operators taking measurement data belonging to a space $\Phi$ to functions belonging to a different space $\Psi$ is under study. This extension seems promising for applications. 

Another present research project concerns the study of the algebraic and topological properties of the spaces of GINOs.

\subsection*{Acknowledgement}

The author thanks Sergio Rajsbaum for his valuable suggestions and advice concerning multiperspectivity. Special thanks to Marc Ethier, Massimo Ferri and Grzegorz Jab{\l}o\'nski for their precious help. The research described in this article has been partially supported by GNSAGA-INdAM (Italy), and is based on the work realized by the author within the ESF-PESC Networking Programme ``Applied and Computational Algebraic Topology''.


\bibliographystyle{eg-alpha-doi}

\bibliography{Patrizio_Frosini_bib}


\end{document}